\documentclass[utf8]{FrontiersinHarvard} 

\usepackage{url,hyperref,lineno,microtype,subcaption,booktabs}
\usepackage[onehalfspacing]{setspace}
\usepackage{comment}
\usepackage{hyperref}   
\usepackage{xr-hyper}   
\usepackage{natbib}

\def\keyFont{\fontsize{8}{11}\helveticabold }
\def\firstAuthorLast{Gangopadhyay,Pessi {et~al.}} 
\def\Authors{Anjasha Gangopadhyay\,$^{1,*}$, Priscila J. Pessi\,$^{1,2,*}$}
\def\Address{$^{1}$ Department of Astronomy, Oskar Klein Center, Stockholm University, SE-106 91 Stockholm, Sweden \\
$^{2}$ Astrophysics Division, National Centre for Nuclear Research, Pasteura 7, 02-093 Warsaw, Poland}
\def\corrAuthor{Corresponding Author}

\def\corrEmail{anjasha.gangopadhyay@astro.su.se, priscila.pessi@ncbj.gov.pl}

\usepackage{tikz}
\usetikzlibrary{arrows.meta,positioning,shapes}

\begin{document}
\onecolumn
\firstpage{1}

\title[Continuity in CCSNe]{Hydrogen-Rich to Stripped-Envelope: Observational Continuity and Biases in CCSNe} 

\author[\firstAuthorLast ]{\Authors} 
\address{} 
\correspondance{} 

\extraAuth{}

\maketitle


\begin{abstract}


Although historically classified into discrete subclasses, there is growing evidence that indicates that core-collapse supernovae (CCSNe) categories often overlap, reflecting continuous variations in progenitor structure, mass-loss history, and circumstellar environments rather than strictly distinct channels. In this review, we explore the proposed continua that link hydrogen-rich Type II SNe to stripped-envelope explosions (IIb$\rightarrow$Ib$\rightarrow$Ic), and that extend further into interaction-dominated and superluminous events. We discuss the physical processes—stellar winds, binary interaction, eruptive outbursts, and circumstellar interaction—that may produce graded outcomes across classes, while highlighting where observational evidence supports or challenges smooth transitions. We propose that CCSNe are better viewed as a multidimensional continuum of explosion outcomes, where traditional subclasses act as reference points rather than strict boundaries. Future progress will rely on large, homogeneous datasets and advanced modeling to disentangle true evolutionary sequences from apparent overlaps, ultimately connecting progenitor pathways to the observed diversity of explosions.

\tiny
 \keyFont{ \section{Keywords:} supernova, photometry, spectroscopy, continuum, photometry, spectroscopy} 
\end{abstract}

\section{Introduction}
Core-collapse supernovae (CCSNe) mark the terminal explosions of massive stars with zero-age main-sequence masses $\gtrsim 8$ M$\odot$ in the case of single-star evolution, and potentially as low as $\sim 5$ M$\odot$ when in a binary system \citep{2021A&A...645A...6Z,Gilkis2025}, whose nuclear burning stages ultimately lead to the formation of an inert iron core. Once the Chandrasekhar limit is exceeded, the core collapses catastrophically, driving densities to nuclear values and forming a proto-neutron star or, in extreme cases, a black hole. The release of gravitational binding energy ($\sim10^{53}$ erg) emerges predominantly as neutrinos, a small fraction of which re-energizes the stalled shock through neutrino heating, aided in many cases by hydrodynamic instabilities such as convection and the standing accretion shock instability \citep[SASI,][]{Bethe1990,Janka2012,Burrows2021}. While the neutrino-driven mechanism is widely accepted as the canonical pathway, rotation, magnetic fields, and fallback accretion may provide additional energy sources, particularly in highly asymmetric explosions or central-engine-powered transients.

The taxonomy of CCSNe originated as an empirical scheme based on observed spectra and light-curve morphology. Early work by \citet{1941PASP...53..224M} distinguished different ``Types'' of events based on spectroscopic similarities. The largest number of events was classified as ``Type~I'', while a fewer number of objects showing hydrogen (H) features was classified as ``Type~II''. \cite{1941PASP...53..224M} also proposed a potential 3rd Type (Type III) composed by a single event that showed sufficiently different spectroscopic features. Later on, 
light curve characterization and further spectroscopic analysis \citep[e.g.][]{Kulikovskii1944,1964AnAp...27..300Z} led to the designation of a 4th and 5th Type (Type IV and V). \cite{1964AnAp...27..300Z} noticed that only the light curves of the Type I events were reasonably consistent while other types had a variety of light curves differences, attributed to asymmetries, the presence of circumstellar material (CSM) or background contamination. Of all the types, only Types I and II survived. Over subsequent decades, this simple division blossomed into a wide family of subclasses—II-P/L, IIn, IIb, Ib, Ic, broad-lined Ic (Ic-BL), and, more recently, interaction-dominated Ibn and Icn—that reflect diverse photometric and spectroscopic features, hinting towards a diversity on their progenitors, circumstellar environments, and explosion conditions. As larger samples became available and theoretical modeling matured, it became increasingly clear that these subclass ``boxes'' are porous: many objects bridge categories, and distributions of luminosity, color, velocity, and spectral line strengths overlap substantially. In this sense, CCSNe could be understood as occupying a \emph{continuum} of outcomes, with apparent subclasses emerging from combinations of envelope mass, chemical composition, pre-supernova (pre-SN) mass loss, and ejecta–CSM coupling, rather than from discrete progenitor channels alone \citep[e.g.,][]{Anderson2014,Lyman2016,Prentice2019,Dessart2020}.

Massive-star evolution proceeds along multiple pathways in which the residual hydrogen (H) and helium (He) at core collapse depend sensitively on metallicity-dependent winds, eruptive or binary-driven mass loss, rotation, and mixing \citep{Laplace2021,Gilkis2025}. Binary interaction is especially influential: Roche-lobe overflow and common-envelope phases efficiently peel away outer layers, yielding a spectrum of final envelope masses from H-rich II, to He-rich Ib, to He-poor Ic \citep[e.g.,][]{Liu2016,Taddia2018,Prentice2019}. The mass stripped prior to explosion sets the outer density/composition profile and, together with $^{56}$Ni mass and explosion energy, governs diffusion timescales and peak luminosity. Meanwhile, the density, composition, and geometry of any CSM laid down by winds or eruptions modulate the emergent observables via shock interaction, adding power and imprinting narrow/intermediate-width emission lines in spectra \citep[flash features, IIn, Ibn, Icn; ][]{2014Natur.509..471G,2016ApJ...818....3K,Hosseinzadeh2017,Fraser2020,Pellegrino2022}. Observationally, He identification in Ib depends on non-thermal excitation, which can blur the Ib/Ic boundary \citep{Hachinger2012,Dessart2020}; similarly, IIb events demonstrate how even a thin residual H envelope dramatically alters early spectra and light curves, bridging Type II and stripped-envelope (SE) SNe. Across this diversity, typical explosion energies cluster around $\sim10^{51}$ erg, ejected masses range from a few to several tens of solar masses, and the synthesized $^{56}$Ni masses span $\sim10^{-3}$ to $0.5$~M$_\odot$ \citep{Hamuy2003,Pejcha2015,Anderson2019}. Their light curves and spectra encode information about progenitor structure, mass-loss history, and circumstellar environments, revealing the central role of stellar winds and binary interactions in shaping the explosion outcomes. In this review we consider all this and discuss the proposed continuity that links H-rich Type~II SN to the SE-SN sequence (IIb$\rightarrow$Ib$\rightarrow$Ic).

\begin{figure}[htbp]
    \captionsetup{aboveskip=3pt, belowskip=-3pt} 
    \centering
    \includegraphics[width=0.9\textwidth]{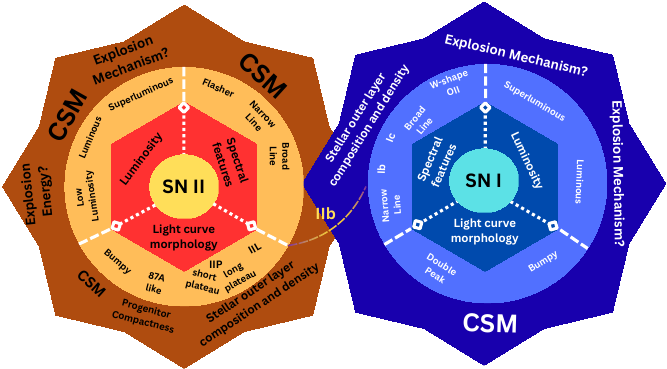}
    \caption{Schematic illustration of the current classification scheme of supernovae based on spectral features, light curve morphology and luminosity.Because spectral classification can overlap with photometric classification (e.g. SLSN~IIn) we refrain from placing such classes, instead we place ``superluminous'' as a class based on luminosity and ``narrow line'' as a class based on spectral lines}.
    \label{fig:sn_classification}
\end{figure}

\section{Classes}

As mentioned above, the presence or absence of certain spectral features determine the primary classification of a SN, however many classes have arisen that further highlight light curve features such as luminosity or overall morphology. Hybrid subclasses can also arise, such as SLSN~IIn \citep[e.g.][]{2018MNRAS.475.1046I} or Luminous SN~I \citep{2022ApJ...941..107G}, which combine characteristic high luminosities with the presence of narrow H features or the complete absence of H, respectively. Interpreting the dominant powering mechanism for each class becomes increasingly uncertain, with scenarios involving CSM interaction, progenitor structure, and energy deposition all remaining viable. As a result, sharp boundaries between classes tend to blur. Fig.~\ref{fig:sn_classification} intends to capture such overlap of different subclasses.

\subsection{Type II SNe}

Type II are the most abundant type of CCSNe \citep{2011MNRAS.412.1441L,2011MNRAS.412.1522S,2020ApJ...904...35P}. Their light curves show a large diversity which was seen very early on, when the first comparative studies were performed on a handful of events (e.g., \citealt{1967SvA....11...63P}), and keeps appearing today, when analyzing samples that are orders of magnitude larger (e.g.,\citealt{2024A&A...692A..95A,2025MNRAS.541..135H}). It becomes natural that the class of H-rich Type II supernovae (SNe~II) is further subdivided based on specific light curve features. 

Type IIP and IIL SNe belong to the oldest subdivision, which separates events based on the shape of the light curve after peak, and assigns the IIP subclass to events that show a ``plateau'', and the IIL subclass to events that decline linearly \citep{1979A&A....72..287B}. Photometrically, they share similar rise times of $\sim$ 7--15 d in optical bands \citep[][]{2015A&A...582A...3G,2015MNRAS.451.2212G,2016ApJ...820...33R,2016MNRAS.459.3939V,2019MNRAS.488.4239P,2025MNRAS.541..135H}. Spectroscopically, they show comparable evolution of the characteristic H$\alpha$ feature \citep{2017ApJ...850...90G}.

The discovery of SN~1987A (\citealt{1989ARA&A..27..629A,2017hsn..book.2181M}), showing a peculiar light curve with a long, slow rise to peak, triggered the creation of the 87A-like subclass, that groups events with similarly long rising light curves. These are spectroscopically similar to SNe~IIP/IIL, but show rise times of $\sim$ 80 days (\citealt{2023ApJ...959..142S}).

Following the discovery of events with unprecedented light curve peak luminosities ($\lesssim -20$~mag in optical bands), the SLSN~II subclass was born (see \citealt{2019ARA&A..57..305G}, for a review). Events that belong to this subclass show a large diversity of spectral and light curve features, although they are typically long lived, taking anywhere from $\sim$ two weeks to more than 80 days to rise, and up to more than a year to decline (\citealt{2022MNRAS.516.1193K,2025A&A...695A.142P}). This luminosity-based distinction from typical IIP/IIL light curves has further motivated the definition of two additional subclasses: low-luminosity SNe~II (LLSNe~II; \citealt{2004MNRAS.347...74P,2014MNRAS.439.2873S,2025PASP..137d4203D}) and luminous SNe~II (LSNe~II; \citealt{2023MNRAS.523.5315P}).

SNe~II are also subdivided by the width of the H$\alpha$ emission. Events showing a narrow, centrally peaked (“Eiffel-tower” like) H$\alpha$ component are classified as SNe~IIn \citep{1990MNRAS.244..269S}. This narrow emission traces ejecta–CSM interaction with H-rich material rather than progenitor composition \citep{2017hsn..book..403S,2021MNRAS.506.4715R}. The persistence of the narrow feature varies widely, and objects may be labeled IIn even if the feature fades ``early'' (i.e. do not persist throughout the entire SN evolution). Some SNe~II exhibit such narrow lines only at very early times ($\lesssim$20\,d) due to flash ionization of nearby CSM, these are known as “flashers” \citep{2014Natur.509..471G,2016ApJ...818....3K,2021ApJ...912...46B,2023ApJ...952..119B}. The distinction between long flashers and SNe IIn with narrow lines that fade early remains unclear. The spectroscopic IIn label is often combined with photometric subclasses such as IIn-P and SLSN~IIn \citep{2013MNRAS.431.2599M,2018MNRAS.475.1046I}.

A subclass exists that exhibits strong H features in its early spectra, which fade as the SN evolves, eventually revealing prominent He lines, similar to those of the SN~Ib class. Thus, this subclass has been dubbed IIb (\citealt{1993ApJ...415L.103F,1996ApJ...459..547C}). Their light curves may be either single- or double-peaked. When present, the first peak arises from emission in the shock-heated hydrogen envelope \citep{2014ApJ...788..193N}, followed by a broader, bell-shaped peak with rise times of $\sim$16–26 days \citep{2019MNRAS.488.4239P}, powered primarily by radioactive decay. The initial peak is often missed, either because the explosion occurs in a low-density environment or due to delayed follow-up observations.
The transition from H-rich to H-poor spectra is attributed to the progenitor having lost a large fraction of its H outer envelope, with only a small amount of H remaining, which has led the community to consider SNe~IIb as part of the SE-SN population (e.g., \citealt{1996ApJ...459..547C,2001AJ....121.1648M}). As such, they are considered to be the link that bridges together the II-IIb-Ib-Ic classes.

\subsection{Continuum between the H-rich SN subclasses}
\label{H-rich-continuum}

A continuum of light curve morphologies is observed across SNe~IIP-IIL \citep{Anderson2014,2015ApJ...799..208S,2016AJ....151...33G,2016ApJ...828..111R}. Modeling distinct SNe light curve phases constrains progenitor and explosion properties \citep[e.g.,][]{2022A&A...660A..40M}. Thus, the photometric continuum plausibly reflects a continuum in progenitor characteristics, most prominently the H-envelope mass \citep{1983Ap&SS..89...89L,1992SvAL...18...43B,1993ApJ...414..712P}. In this picture, larger H envelopes produce longer plateaus; as the available H for recombination decreases, the plateau shortens and transitions toward a linear decline. The continuum extends to optical spectroscopy, where several spectral parameters correlate with photometric ones \citep{2017ApJ...850...90G}, although such continuum seems to not extend to the NIR \citep{2019ApJ...887....4D}, where distinct behavior is seen primarily in the He I features of corresponding subclasses.

It has been proposed that progressive stripping of the H envelope could link SNe~IIP/IIL to IIb, with the weakening and eventual disappearance of H features in SNe~IIb reflecting a decreasing H mass in the progenitor, i.e., a putative IIP-IIL-IIb sequence. Binary-population and explosion models can indeed yield quasi-continuous outcomes \citep{2018PASA...35...49E,2024A&A...685A.169D,Gilkis2025}. However, observations do not support a smooth continuum: IIP/IIL and IIb exhibit distinct light-curve properties \citep{2012ApJ...756L..30A,2019MNRAS.488.4239P,2023MNRAS.518.5741S} and spectroscopic behavior \citep[more noticeable $\sim$20\,days after explosion][]{2025arXiv250708731G}.

Stripping of the progenitor’s outer H layers can contribute to the CSM, raising the question of where SNe~IIn sit within a continuum of H-rich SNe. If the CSM is primarily built by pre-SN mass loss, a sequence with increasing mass loss is expected: (i) low mass loss yields regular SNe~IIP/IIL; (ii) intermediate mass loss produces more luminous SNe~II with broader, ``boxy'' H$\alpha$ features \citep[similar to those in][]{2023MNRAS.523.5315P} resulting from the expansion of a shell formed by the reverse shock from the ejecta-CSM interaction \citep{1995A&A...299..715P,2016MNRAS.456.1269B,2022A&A...660L...9D}; and (iii) high mass loss produces either a progenitor with not enough H for the resulting SN to show H features throughout its entire evolution, or a dense CSM that is optically thick to electron scattering and generates narrow emission lines when ionized by the SN ejecta, while also producing more luminous light curves through the conversion of kinetic energy into radiation at the radiative shock \citep[e.g.,][]{2017hsn..book..403S}. 
Thus we could think of two possible continua, one that considers SNe IIP--IIL--IIb and one that considers SNe IIP-IIL-LII-IIn/SLSN~II. If both these continua are linked by progenitor mass/mass loss, then some intermediate ``SN~IIbn'' should exist, where the interaction is evidenced by narrow lines throughout the SN evolution, but where prominent H lines disappear to give rise to prominent He lines. Some evidence exist hinting possible common features between SN~IIb and IIn. Narrow flash ionization features indicative of CSM presence have been detected in SN~IIb \citep[e.g.,][]{2016ApJ...818....3K}. Yet, precursor outbursts associated with pre-explosion stellar mass loss \citep{2010AJ....139.1451S,2022ApJ...936..114M,2023ApJ...945..104T}, common among SNe~IIn \citep{2014ApJ...789..104O,2021ApJ...907...99S,2024A&A...686A.231R}, appear to be rare in SNe~IIb \citep{2015ApJ...811..117S}. This suggests that, while both SNe~IIn and SNe~IIb progenitors undergo significant envelope loss, the underlying mass-loss mechanisms may differ. It has been proposed that some SNe~Ibn may originate from a SN~IIb explosion embedded in a dense CSM \citep{2020MNRAS.499.1450P}. In such cases, the transient initially exhibit SN~IIb-like features and subsequently evolve into a SN~Ibn as the ejecta interact with He-rich CSM. This scenario implies that a subset of SN~IIb progenitors may undergo not only extensive H envelope stripping but also substantial He loss prior to core collapse.

Some events appear to bridge the Type II and IIb classes, e.g: SN~2013ai \citep{2021ApJ...909..145D}; SN~2017ivv \citep{2020MNRAS.499..974G} and; SN~2018gk \citep{2021MNRAS.503.3472B}. SN~2013ai is spectroscopically a type II, but the velocity and strength of different features is more similar to those of stripped envelope SN, in addition its light curve rise time is significantly larger than typical SN~II. SN~2017ivv and SN~2018gk are also considered to belong to the IIb class based on their broad H$\alpha$ components and on the appearance of late time spectral features. Some transitional events have been proposed to link the IIb-Ib classes. For example SN~2022crv, which showed early H that vanished quickly, and SN~2020cpg, which exhibited H features over otherwise SNe~Ib characteristics \citep{Dong2024,Gangopadhyay2023,2022MNRAS.517.5678T,Medler2021}. It has been shown that a fraction of SNe~Ib exhibit weak H features in their spectra. These events are not classified as SNe~IIb but rather support the existence of a spectral continuum between the Ib-IIb classes, evident in both the optical and NIR regimes \citep{Liu2016,2022ApJ...925..175S}.

\begin{table*}[ht!]
\centering
\caption{Summary of light-curve and spectral diagnostics for hydrogen-rich CCSNe.\label{tab:ccI_summary}}
\footnotesize
\setlength{\tabcolsep}{3pt} 
\begin{tabular}{@{}lccccp{6.0cm}p{1.5cm}@{}}
\toprule
Subtype & Rise to peak (d) & Peak M$_{V}$ (mag) & $M_{^{56}\mathrm{Ni}}$ (M$_\odot$)   & Peak $v_{\rm ph}$ (km s$^{-1}$)  & Hallmark spectral features  \\
\midrule
IIP/L    & 7--15      & $-14.0$ to $-18.5$  & 0.001--0.36              & 5\,000–12\,000                &    Balmer lines (P-Cygni H$\alpha$)                      \\
87A-like & 70--140    & $-15.6$ to $-17.5$  & 0.04--0.47               & 6\,000–15\,000                &    Balmer lines (P-Cygni H$\alpha$)                      \\
IIn      & 12--135    & $-16.8$ to $-20.0$  & 0.007--0.4 / CSM-dominated & 5\,00–1\,000    &   Balmer lines (narrow H$\alpha$) \\
SLSN~II  & 15--100    & $-19.9$ to $-22.0$  & 0.04--8.4 / CSM-dominated    & 9\,00–12\,000  &    Balmer lines (P-Cygni or broad or narrow  H$\alpha$) \\
IIb      & 16--26     & $-16.1$ to $-18.7$  & 0.03–0.28                & 5\,000–10\,000                &   Balmer lines (H disappear to He I lines) \\
\midrule
\midrule
Ib    & 10--20  & $-16.5$ to $-18.0$          & 0.05--0.2           & 8\,000–12\,000 & He P-cygni lines \\
Ic    & 10--20  & $-16.5$ to $-18.0$          & 0.05--0.2           & 9\,000–12\,000  & Si P-cygni lines\\
Ic-BL & 8--15   & $-17.5$ to $-19.5$          & 0.2--0.5            & 15\,000–30\,000 & Broad Si lines \\
Ibn   & few--10 & $-17.0$ to $-19.0$  & CSM-dominated & 1\,000–5\,000$^\dagger$ & Narrow/IW He\,\textsc{i}, weak/absent H\\
Icn & few--10 & $-17.0$ to $-19.0$  & CSM-dominated & $\cdots$ & narrow/intermediate C\,\textsc{ii/iii/iv}, H/He-poor CSM lines \\
\bottomrule
\end{tabular}
\raggedright
\textit{Notes.} Ranges are approximated and indicative from population studies; we exclude LSNe~II from the table as further studies are needed to constrain their parameters, individual objects span wider diversity. Nickel masses and velocities for IIn/SLSN~II and Ibn/Icn are often ill-defined from diffusion-phase modeling, extreme values are unphysical because alternative powering mechanisms such as interaction dominates. $v_{H_{\alpha}}$ for IIn/SLSN~II are often ill-defined as narrow lines are broadened by electron scattering.$^\dagger$Line widths quoted for the narrow/intermediate He\,\textsc{i} components characteristic of CSM interaction. Representative sources for Type II SNe: \citet{2013A&A...555A..10T,Anderson2014,2014AJ....147..118R,2017ApJ...850...89G,2018MNRAS.475.1046I,2019MNRAS.488.4239P,Anderson2019,2020A&A...637A..73N,2022MNRAS.516.1193K,2023MNRAS.521.4801P,2023ApJ...959..142S,2025A&A...695A..29S,2025A&A...695A.142P,2025arXiv250708731G}. Representative for Type I SNe sources: \citet{Lyman2016,Taddia2018,Prentice2019,Liu2016,Taddia2019,Cano2017,Hosseinzadeh2017,Fraser2020,Hosseinzadeh2019,Pellegrino2022,Gangopadhyay2025}.
\end{table*}

\subsection{Type I SNe}
\label{subsec:snI}

H-poor CCSNe—Types Ib, Ic, Ic-BL, SLSN-I and the interaction-dominated Ibn/Icn—are traditionally grouped as ``Type~I" CCSNe. 
Their photometric and spectroscopic properties reveal both overlaps and distinctions.

Type~Ib SNe are characterized by prominent He\,\textsc{i} lines 
, while Type~Ic lack H and obvious He features, instead showing strong Si and O\,\textsc{i}.
Observations and detailed modeling demonstrates that small amounts of H and He can exist in SNe~Ib and SN~Ic, respectively, although it can
remain undetected 
\citep{Hachinger2012,Liu2016,Prentice2019,Dessart2020}, implying a possible continuum in envelope stripping among these classes. Photometrically, SNe~Ib and SNe~Ic share similar rise times (10–20~d) and peak magnitudes ($M_V\!\sim\!-16.5$ to $-18$ mag), with overlapping ejecta and $^{56}$Ni masses \citep{Lyman2016}. 

Broad-lined Ic (Ic-BL) represent the high-velocity tail, with $v_{\rm ph}\gtrsim1.5\times10^4$ km s$^{-1}$, high kinetic energies ($\gtrsim10^{52}$ erg), and frequent associations with long-duration gamma-ray bursts \citep{Cano2017,Taddia2019}. They demonstrate the extreme of the SE-SNe. 

At the higher luminosity end of the SNe~Ic, lie a group of members with peak absolute magnitudes of $\lesssim -$21  mag defined as Type I superluminous SNe (SLSNe~I—often termed SLSNe~Ic). They often show hot, blue spectra dominated by broad C/O features \citep{Quimby2011,Nicholl2015,Lunnan2018,Yan2017}. Their light curves show wide diversity in rise times (from weeks to $\gtrsim100$\,d) and frequently exhibit an early “precursor” bump \citep{Inserra2013,Leloudas2012,Nicholl2016,Lunnan2016}. Post-peak, SLSNe~Ic split into fast and slow decliners; while some slow events roughly track the $^{56}$Co decay rate, additional power (e.g., a central engine and/or CSM interaction) is generally required \citep{Nicholl2015,Gal-Yam2009,Gal-Yam2012,Moriya2017}.

Interacting H-poor classes
are primarily governed by the composition of the CSM into which the ejecta expands.
SNe~Ibn show narrow/intermediate-width He emission lines, rapid blue light curves, and evidence for dense He-rich CSM, with typical mass-loss rates of $\sim10^{-2}$–0.1 M$_\odot$ yr$^{-1}$, possibly associated to Wolf--Rayet (WR) or WR-like stars \citep{Pastorello2008,Hosseinzadeh2017,Fraser2020}. Recently recognized SNe~Icn are dominated by narrow/intermediate C lines and very blue early emission, signaling explosions within dense, C-rich CSM, likely linked to progenitors stripped even further of their outer layers \citep{Gal-Yam2022,Pellegrino2022}. These events demonstrate that extreme late-stage mass loss can strip progenitors nearly free of H and He, yielding C/O-rich winds just prior to collapse.

Table~\ref{tab:ccI_summary} summarizes different parameters for CCSNe, and shows that Type I subclasses form overlapping distributions in rise times, luminosities, Ni masses, velocities, and spectral diagnostics. Environmental studies show systematic trends: Ic occur preferentially at higher metallicities than Ib, consistent with stronger winds, while Ibn/Icn trace environments of massive WR-like progenitors \citep{Sanders2012,Fremling2018}. Together, these findings indicate that Type~I CCSNe can represent a continuum of stripping outcomes, with Ib and Ic linked through varying He retention and with Ibn/Icn reflecting the most extreme CSM-loaded progenitors.

\begin{figure}[htbp]
    \centering
    \includegraphics[width=\textwidth]{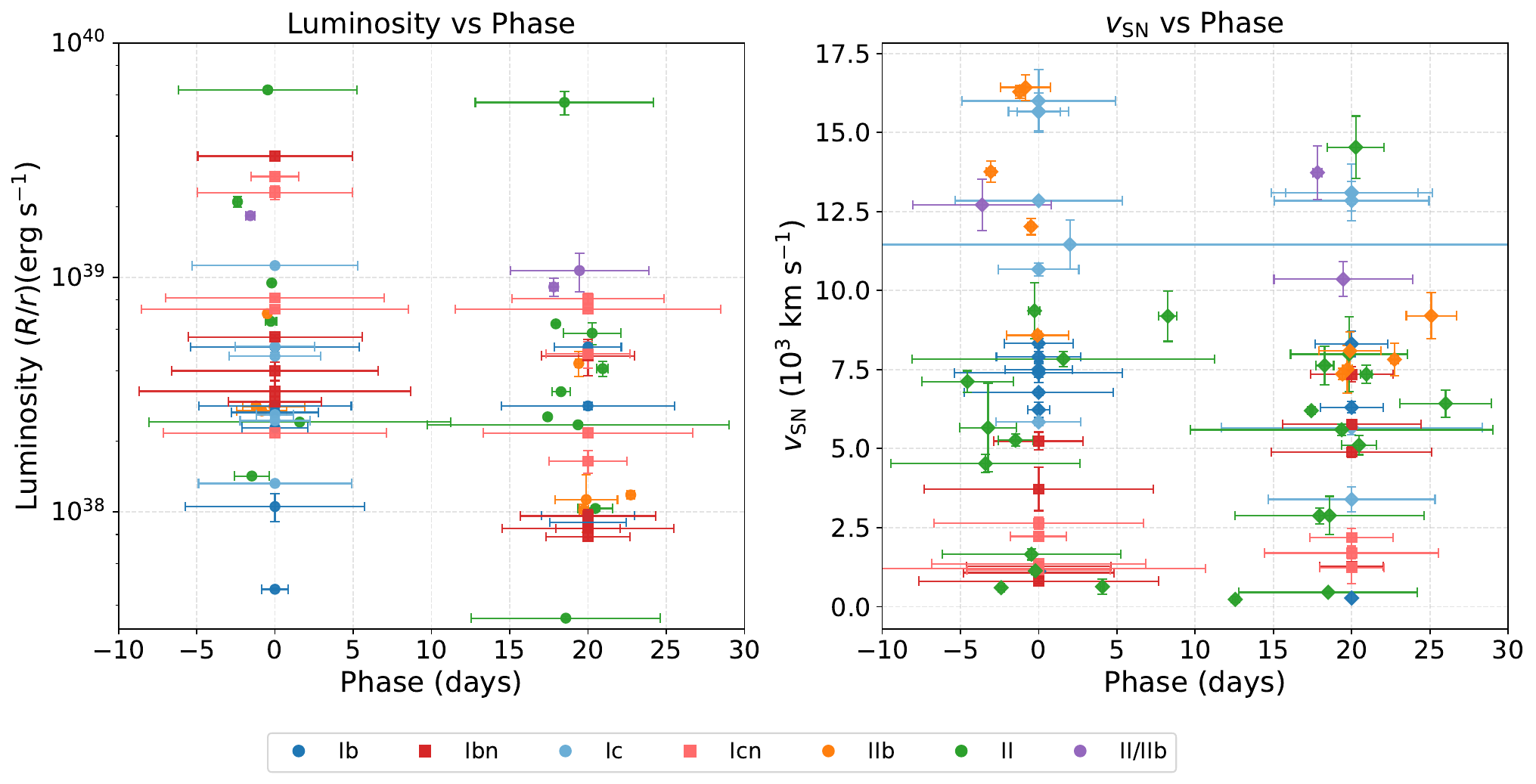}
    \caption{Luminosity $L$ and expansion velocity $v_{\rm SN}$ as functions of phase.
    CCSNe spread all over the phase space in the luminosities for different subtypes while for velocities, the velocity has higher value for SESNe than the interacting SNe as the latter are dominated by electron scattering wings. Error bars denote $1\sigma$ uncertainties. Note that SNe~II include a large variety of subclasses including SNe~IIn (see Sect.\ref{supp:lumvel}). Different markers are used to separate the interacting and the non-interacting classes. Cautionary note: velocity tracers for interacting SNe are not directly comparable to photospheric velocities of non-interacting SNe. Measured velocities for interacting events (derived from narrow or intermediate-width lines) reflect slow CSM or post-shock gas, rather than ejecta photospheric speeds. Consequently, comparisons of ($v_{\text{SN}}$) across classes are not strictly homogeneous. The apparent lack of a clear continuum in the plotted luminosity–velocity distribution may partly reflect measurement bias, rather than disproving an underlying physical continuum.}
    \label{fig:LvV}
\end{figure}

Fig.~\ref{fig:LvV} was created using publicly available data collected from dedicated articles, the Open Supernova Catalog\footnote{Although the Open Supernova Catalog is no longer actively maintained, its data remain accessible in the repository at \url{https://github.com/astrocatalogs/supernovae}.} \citep{2017ApJ...835...64G} and/or WiseRep \citep{wiserep2012}.
The sample consists of: SNe~II (including IIP, IIL, LSN~II, SLSN~II, IIn, LLSN~II): 1987A \citep{1990AJ.....99.1146H,1995ApJS...99..223P}; 1998S \citep{2000MNRAS.318.1093F,2001MNRAS.325..907F,2005ApJ...622..991F}; 1999em \citep{2014MNRAS.442..844F,2016AJ....151...33G,2001ApJ...558..615H,2002PASP..114...35L}; 2006Y, 2006ai, 2008bk \citep{Anderson2014,2017ApJ...850...89G,2024A&A...692A..95A}; 2008es \citep{2009ApJ...690.1313G,2014MNRAS.445..554F}; 2009aj \citep{2017ApJ...850...89G,2020MNRAS.494.5882R,2024A&A...692A..95A}; 2014G \citep{2019MNRAS.490.2799D,2016MNRAS.462..137T}; iPTF14hls \citep{2017Natur.551..210A,2019A&A...621A..30S}; 2021acya \citep{2025A&A...695A..29S}; 2022lxg \citep{2025A&A...700A.138C};\\
SNe~II/IIb: 2013ai \citep{2021ApJ...909..145D,2016PASA...33...55C}; 2018gk \citep{2021MNRAS.503.3472B};\\
SNe~IIb: 1993J \citep{1995A&AS..110..513B,1996AJ....112..732R}; 2006T \citep{2014ApJS..213...19B,2023A&A...675A..82S}; 2008ax \citep{2008MNRAS.389..955P,2009PZ.....29....2T,2011MNRAS.413.2140T}; 2011dh \citep{2014A&A...562A..17E}; 2016gkg \citep{2017ApJ...836L..12T};\\
SNe~Ib: 1999dn \citep{Benetti2011}; 1990I \citep{Elmhamdi2004}; 2007Y \citep{Stritzinger2009}; 2009jf \citep{Valenti2011}; iPTF13bvn \citep{Srivastav2009jf}. \\
SNe~Ic: 1994I \citep{Richmond1996}; 2002ap \citep{Foley2003}; 2004aw \citep{Taubenberger2006}; 2007gr \citep{Hunter2009}; 2016coi \citep{Kumar2018};\\ SNe~Ibn: 2006jc \citep{Pastorello2008}; 2010al \citep{Pastorello2015}; 2019uo \citep{Gangopadhyay2020}; 2019wep \citep{Gangopadhyay2022}; ASASSN-15ed \citep{Pastorello2015b};\\
SNe~Icn: 2019hgp \citep{Gal-Yam2022}; 2019jc \citep{Pellegrino2022}; 2021ckj \citep{Pellegrino2022}; 2021csp \citep{Pellegrino2022,Perley2022}; 2022ann \citep{Davis2023}.\\
For our CCSN sample, we estimated the $R$/$r$-band luminosity at peak and 20 days post-peak by interpolating the light curves with Gaussian Processes using the \texttt{GPy} Python package \citep{gpy2014}. At the same epochs, we measured the full-width half-maximum velocities of key emission lines: Hydrogen (Type II), Helium (Type Ib, Ibn), Silicon (Type Ic), and Carbon (Type Icn). To this end, corresponding emission features were fit with Gaussians profiles using the \texttt{lmfit} Python package \citep{2014zndo.....11813N}.

\subsection{Continuum between the stripped envelope SN subclasses}

Although H is present in the spectra of SNe~IIb, they are still classified within the SE-SN group. Extensive spectroscopic and photometric samples show that SE-SNe—IIb, Ib, and Ic—represent a continuum of envelope stripping: SNe~IIb retain a small amount of H, SNe~Ib arise from He-rich but H-poor progenitors, and SNe~Ic are produced by the most heavily stripped progenitors \citep{Liu2016,Taddia2018,Prentice2019}. 
Systematic trends have been found in He and O line strengths and in environmental metallicities that track the stripping sequence. Two opposing views suggest that either Ic are not hiding He but are genuinely more stripped than Ib \citep{Fremling2018,Shivvers2019} or, that some SNe~Ic may have a significant fraction of their He that is effectively transparent \citep{Piro2014}. Evidence of the former is that the strength and characteristic velocity of the O 7774 \AA~ absorption feature are higher in SNe Ic than in SNe Ib and IIb at phases near peak brightness highlighting a more stripped progenitor for SNe Ic than IIb/Ib. Evidence of the latter lies in theoretical modeling that indicates relatively low velocities with little velocity evolution, as are expected deep inside an exploding star, along with temperatures that are too low to ionize He, but no detailed observational testing has been done to validate such theoretical framework.
\cite{10.1093/mnras/stx980} through the pseudo-equivalent width measurements of SNe~IIb/Ib lines have deciphered a continuity in the outer envelope and sub-classified them as IIb, IIb(I), Ib(II) and Ib, which represent H-rich to H-poor SNe.
Detailed spectral analyses using recent Machine Learning algorithms also show weak high-velocity H in some Ib and hints of He in rare Ic, consistent with a continuum rather than sharp subtype divisions \citep{Holmbo2023}. Together, these studies argue that SE-SNe could form a spectroscopic and photometric continuum reflecting the extent of mass stripping prior to explosion. However, SNe~Ic could be a fundamentally distinct category, rather than SNe~IIb/Ib which are more linked by continuum of outer H/He envelope.

Observational evidence also supports a continuity between the interacting subclasses. The relative strengths of He, and C/O features provide a natural way to place events along the Ibn-Icn sequence. 
WR-stars of the WN sequence, with He-dominated winds, are natural progenitors of SNe~Ibn, while WC/WO stars with C/O-rich winds provide a plausible channel for SNe~Icn. Transitional events such as SN~2023xgo, which showed mixed He and stronger C features early on (\citealt{Gangopadhyay2025b}), suggest that Ibn and Icn are not entirely disjoint, but part of a continuous sequence tracing increasingly stripped progenitors. In this framework, SNe~Ibn and Icn can be viewed as adjacent points along the same continuum of ejecta--CSM interaction, where the observational subclass reflects the chemical composition of the last layers shed by the star prior to explosion.

There is limited phenomenological continuity between SLSNe~Ic and normal Ic. Spectroscopically, SLSNe~Ic often evolve to Ic/Ic-BL-like appearances at late times with comparable photospheric velocities \citep{Pastorello2010,Nicholl2016}, and “luminous Ic”/Ic-BL events partly populate the intermediate luminosity regime, hinting at a bridge \citep{Arcavi2016,Roy2016}. However, population studies indicate a bimodal luminosity function—Ic peaking near $M_{\rm bol}\sim -17.5$ to $-18.5$ \citep{Taddia2015,Lyman2016} and SLSNe~Ic clustered around $M_{\rm bol}\sim -21\pm0.5$ \citep{Nicholl2015}—with in between objects seemingly bridging the Ic-SLSN-I classes \citep{2022ApJ...941..107G}. Moreover, SLSNe~Ic generally require engine/interaction power beyond $^{56}$Ni \citep{Gal-Yam2012,Moriya2017}. Any ``continuum" thus likely reflects varying engine strength and/or CSM interaction rather than a smooth extension of the Ic population in nickel yield. Some recent cases of SLSN~I have some signatures of late-time CSM interaction with previously expelled shells which can be H-rich \citep{Pursiainen2022}. Some recent cases of SLSNe~I such as 2018bsz \citep{2018A&A...620A..67A,Pursiainen2022}, PTF13ehe (H$\alpha$ emerging at $\geq$ 200–300 d), iPTF15esb and iPTF16bad \citep{Yan2015,Yan2017} have shown late time interaction signatures which can probe mass-loss history of these SNe. 

\section{Looking into the luminosity and velocity of core-collapse SNe}
\label{luminosity-velocity}

We map the luminosity–velocity (\(L\)–\(v_{\rm SN}\)) space for a heterogeneous sample of CCSNe; the sample definition and references are given in Supplementary Material (Sect~\ref{supp:lumvel}).  Our initial goal was to estimate mass-loss rates to trace the pre-SN progenitor history. However, the assumption that shock power alone drives the luminosity of these SNe does not hold across all CCSN sub-classes. We therefore adopt a model-agnostic comparison of the observables that enter mass-loss prescriptions (e.g., \citealt{Chugai1994}): a bolometric luminosity and SN velocity proxy. Specifically, we use the \(R/r\)-band flux as a tracer of \(L\) and full-width half-maximum velocity of characteristic emission lines (H/He/Si/C) as SN velocity.

Figure~\ref{fig:LvV} shows that across $-10$ to $+30$\,d, CCSNe populate a largely continuous locus in luminosity $L$ and line velocity $v_{\rm SN}$, with substantial overlap between different groups of CCSNe. At both $\sim$0 and $\sim$20\,d, the \emph{monochromatic} $R/r$-band luminosities do not cleanly separate interacting from non–interacting events and are fairly distributed all over the phase space. Ideally, we would expect the interacting SNe to be more luminous than their non-interacting counterparts as interaction drives as an additional luminosity powering source and not only radioactivity. The observed behaviour in Fig.~\ref{fig:LvV} is justified as interaction power is often emitted at bluer/UV and high–energy wavelengths; a single red optical band, therefore do not acount for the full bolometric output and is highly sensitive to temperature, line blanketing, and extinction. The distribution further indicates that radioactive heating is not the sole luminosity source in many CCSNe and that additional power source contributes with phase and band-dependent visibility.

In velocity space, objects with clear interaction signatures tend to show lower measured $v_{\rm SN}$ than non–interacting SNe. This reflects a \emph{measurement–tracer effect}: for interacting SNe, the narrow/intermediate lines (with possible electron–scattering wings) predominantly trace slowly moving, pre–shock CSM or post–shock gas rather than the ejecta photosphere, whereas non–interacting SNe are measured from intrinsically broad P-Cygni ejecta features. Consequently, the quoted $v_{\rm SN}$ values are not strictly homologous across sub–classes.

\section{Discussion}

This review discusses a continuity framework for CCSNe, wherein the traditional subclasses are best understood as overlapping regions in a small set of physically meaningful dimensions: residual H/He in the envelope, $^{56}$Ni mixing and heating, ejecta kinetic energy, and the density/composition of CSM. In this view, the familiar labels (II, IIb, Ib, Ic, Ic-BL; and the interacting IIn/Ibn/Icn) are not discrete boxes but landmarks along graded paths set by mass loss. 
Constraining the mass-loss history of a progenitor requires two readily observable parameters: the bolometric luminosity and the SN velocity at a given epoch (see Sect.~\ref{luminosity-velocity}), the latter typically estimated from the full-width half-maximum of characteristic spectral lines. In practice, however, true bolometric light curves are in most cases unavailable, and emission-line widths reflect not only bulk expansion but also additional broadening mechanisms such as electron scattering, optical depth effects, and radiative transfer. These factors complicate a direct interpretation of observables in terms of progenitor mass loss. Still, if interaction was the dominant contributor to the powering of the SNe, some trends could be expected in the luminosity and velocity distributions. In Fig.~\ref{fig:LvV} we do not see clear continuum among different CCSNe classes (see Sect.~\ref{luminosity-velocity}). 

The literature has often invoked continuous distributions of various parameters to support the idea of a continuum across CCSN (see Table~\ref{tab:classcontinua} for a summary). However, some of these claimed continua can be challenged. For example, it has been suggested that the main driver of SNe~IIP/IIL light curve diversity may be the explosion energy \citep{2009ApJ...703.2205K,2022A&A...660A..42M}. Moreover, \cite{2025arXiv250811077B} argue that the IIP/IIL morphologies arise from progenitors exploding in a compressed/expanded phase of a stellar pulsation. It is also still debated whether a continuum exists within the SN~IIn class itself as multimodality has been observed in the decline rate of their light curves \citep{2020A&A...637A..73N,2025ApJ...987...13R} and on their radiated energies \citep{2024arXiv241107287H}. Furthermore, although the continuum IIP-IIL-IIb is recovered by theoretical models, such continuum is not seen in the observed features, with only a small number of Type IIb events occupying the distributions of observational features of Type II events and viceversa (see Sect.~\ref{H-rich-continuum}). It seems like the conditions needed to produce a transitional II/IIb event in this case are much rarer than those necessary to produce ``typical'' events. On a similar note, it has been traditionally defined that the SESNe group of Type IIb, Ib, Ic should be linked by a continuum of outer envelope \citep{Taddia2018}. While \cite{Prentice2019} find that SNe~IIb/Ib are linked by an outer envelope promoting further sub-classification through pseudo-equivalent width measurements of different sub-classes, extending this sequence to Ic is less secure. Population studies indicate that many SNe~Ic are not merely Ib with ``hidden He’’; instead, their spectra favor genuinely more heavily stripped progenitors, as suggested by systematically stronger and higher–velocity O\,\textsc{i}~$\lambda7774$ and related oxygen diagnostics \citep{Fremling2018,Shivvers2019,Taddia2018}. Thus, the transparent–helium hypothesis remains unproven observationally. Similarly for the SNe~Ibn/Icn, certain events like SN~2023xgo indicates that the CSM around them is composed of both He and C \citep{Gangopadhyay2025}, also implying similar progenitor channels of WR stars or low mass stars in binary companions \citep{Moriya2025,Dessart2022} hinting towards a continuum. 

\begin{table}[!h]
\centering
\renewcommand{\arraystretch}{1.25}    
\setlength{\tabcolsep}{8pt}           
\caption{\textbf{Summary of proposed class continua.}}
\label{tab:classcontinua}
\begin{tabular}{|c|l|l|}
\hline
Class Continua & Support & Evidence \\ \hline\hline
SN~IIP -- SN~IIL  & Well-supported & Observations, models \\ \hline
SN~IIP -- SN~IIL -- SN~IIb & Moderately-supported & Observations, models \\ \hline
SN~IIb -- SN~Ib -- SN~Ic  & Well-supported & Observations, models \\ \hline
SN~II -- LSN~II -- SLSN~II  & -- & No detailed study \\ \hline
SN~II -- SN~IIn  & Weakly-supported & Models \\ \hline
SN~IIn -- SLSN~IIn  & Weakly-supported & Observations \\ \hline
SN~Ic -- LSN~I -- SLSN~I  & Weakly-supported & Models \\ \hline
SN~IIn -- SN~Ibn -- SN~Icn & Well-supported & Observations, models \\ \hline
SN~Ic -- SN~Icn -- SN~Ib -- SN~Ibn & Moderately-supported & Observations, models \\ \hline
\end{tabular}
\end{table}

This raises some fundamental questions: what is the physical significance of observing a continuum in supernova properties? Does a continuum in observables necessarily imply a corresponding continuum in progenitor characteristics and explosion mechanisms? While theoretical models can reproduce a wide range of outcomes, it remains uncertain whether all such outcomes are realized in nature. Theoretical modeling of CCSNe has advanced significantly in recent years, yet most explosion simulations remain restricted to one-dimensional treatments. Large uncertainties also persist from the stellar evolution side, particularly regarding the roles of binarity, metallicity, rotation, magnetic fields, and the relative importance of different mass-loss mechanisms across progenitors. It has been argued that progenitor mass alone does not uniquely determine the fate of a star, with the concept of ``islands of explodability'' emerging from modern simulations \citep{2016ApJ...818..124E,2016ApJ...821...38S}, suggesting that core collapse may occupy a discrete rather than continuous parameter space. Moreover, alternative explosion channels are expected at the extremes of the progenitor mass range: electron-capture SN, triggered by Ne-O deflagration, in the low-mass regime \citep[e.g.,][]{2017hsn..book..483N}; and pair-instability SN, driven by thermonuclear runaway following $e^+e^-$ pair production, in the high-mass regime \citep[e.g.,][]{2011ApJ...734..102K,2024arXiv240716113R}. While neither has yet been observationally confirmed with certainty, both may contaminate CCSN samples. 

Multiwavelength time-domain surveys (and their early high-cadence follow-up) will greatly expand uniform samples. Coupling these data with binary-population synthesis, radiative-transfer tools and 3D modeling, should turn the qualitative ``continuum'' into predictive, multi-parameter maps linking progenitor pathways to observables. 
Such efforts will clarify which apparent continua in CCSN observables reflect genuine physical sequences and which arise from selection effects, degeneracies, or distinct progenitor channels. Ultimately, a unified framework will require mapping the multi-parameter space of mass loss, binarity, CSM structure, and explosion physics onto the rich phenomenology revealed by modern time-domain surveys. As larger and more homogeneous samples accumulate in the coming decade, the field is well positioned to transform the currently qualitative notion of “continuity” in CCSNe into a quantitatively testable, predictive paradigm.

\section*{Conflict of Interest Statement}

The authors declare that the research was conducted in the absence of any commercial or financial relationships that could be construed as a potential conflict of interest.

\section*{Author Contributions}

The authors contributed equally to this work in every role, with each bringing their respective expertise to the study.

\section*{Funding}
A.G is supported by the research project
grant “Understanding the Dynamic Universe” funded by the
Knut and Alice Wallenberg under Dnr KAW 2018.0067. 
P.J.P is funded by the European Union (ERC, project number 101042299, TransPIre). Views and opinions expressed are however those of the author(s) only and do not necessarily reflect those of the European Union or the European Research Council Executive Agency. Neither the European Union nor the granting authority can be held responsible for them.

\section*{Acknowledgments}
P.J.P thanks Laureano Martinez, Claudia Guti\'errez, Ragnhild Lunnan and the OKC supernova group for useful discussion. A.G thanks all the OKC supernova group members for useful discussions.


\section*{Data Availability Statement}
All the data used in this work is public either is the corresponding cited studies, the Transient Name Server, NED and/or WiseRep.

\bibliographystyle{Frontiers-Harvard} 
\bibliography{test}

\end{document}